\documentclass[twocolumn,preprintnumbers,amsmath,amssymb,superscriptaddress]{revtex4}
\usepackage{graphicx}% Include figure files
\usepackage{dcolumn}% Align table columns on decimal point
\usepackage{bm}% bold math
\usepackage{longtable}% bold math
\usepackage{comment}
\usepackage[normalem]{ulem}
\newcommand{\bea}{\begin{eqnarray}}
\newcommand{\eea}{\end{eqnarray}}

\usepackage{amssymb,amsfonts,amsmath,graphics,graphicx}

\begin{document}

%%%%%%%%%%%%%%%%%%%%%%%%%%%%%%

\title{An analysis of a large dataset on immigrant integration in Spain.\\ The Statistical Mechanics perspective on Social Action.}
%The Statistical Mechanics of Social Action in Immigrant Integration: an Empirical Analysis of a Large Dataset in Spain.}
\author{Adriano Barra}
\affiliation{Dipartimento di Fisica, Sapienza Universit\`a di Roma}
\author{Pierluigi Contucci$^*$}
\affiliation{Dipartimento di Matematica, Universit\`a di Bologna}
\author{Rickard Sandell}
\affiliation{Departamento de Ciencias Sociales, Universidad de Carlos III de Madrid}
\author{Cecilia Vernia}
\affiliation{Dipartimento di Scienze Fisiche Informatiche e Matematiche, Universit\`a di Modena e Reggio Emilia}
%\vskip 10truecm
%\date{\today}

%% The \maketitle command is necessary to build the title page.
%\maketitle
%%%%%%
%\newpage
%
%\maketitle
%

\begin{abstract}
\vskip .7truecm
How does immigrant integration in a country change with immigration density? Guided by a statistical mechanics perspective we propose a novel approach to this problem. The analysis focuses on classical integration quantifiers such as the percentage of jobs (temporary and permanent) given to immigrants, mixed marriages, and newborns with parents of mixed origin. We find that the average values of different quantifiers may exhibit either linear or non-linear growth on immigrant density and we suggest that {\it social action}, a concept identified by Max Weber, causes the observed non-linearity. Using the statistical mechanics notion of interaction to quantitatively emulate social action, a unified mathematical model for integration is proposed and it is shown to explain both growth behaviors observed. The linear theory instead, ignoring the possibility of interaction effects would underestimate the quantifiers up to $30\%$ when immigrant densities are low, and overestimate them as much when densities are high. The capacity to quantitatively isolate different types of integration mechanisms makes our framework a suitable tool in the quest for more efficient integration policies.
\end{abstract}


\maketitle

\vskip 1cm

The United Nations recently reported that there are about one billion migrants worldwide, of which one quarter are international migrants \cite{eurep}. The size of the migration phenomenon and the speed by which it increases have turned migration into a challenge that is at the top of the political agenda in the European Union, the United States, and in many other countries across the world. One reason why migration has become a major political priority is that it is a catalyst for large-scale social, economic, and demographic changes \cite{dial} capable of producing opportunities but also turmoil and friction.

{\it Integration} and {\it social cohesion} are keywords when addressing many of the challenges posed by increasing migration. Some of the problems related to such processes are carefully analyzed in several studies (for instance \cite{esrc} and the references therein). The European Union in particular has identified a list of common basic principles to make integration work \cite{eucbp} \cite{eucbp1} based on employment for immigrant, frequent interactions between immigrants and natives, etc.

Our work stems from the simple observation that very little is known about the mechanisms that bring about integration. For example, elementary questions like how integration responds to an increase in immigration density or to what extent the intensity of interaction modifies the level of integration still beg coherent empirical and theoretical answers. Our study is inspired by the realization that precise answers to those questions are of paramount importance to formulating social policies able to promote integration.

The research reported here addresses these issues from a quantitative point of view without making use of the classical historic series approach
(see the Supplementary Material for discussion). Instead we study the problem of integration from the hard-science point of view by relying on methods and techniques stemmed from statistical mechanics, the branch of theoretical physics devised to explain thermodynamic laws as emerging average behavior for systems composed of a high number of microscopic interacting elements. The application of ideas and methods from statistical mechanics to fields other than physics has emerged in several contexts over the past decades (see \cite{mpv}, \cite{st1}, \cite{st2}, \cite{bouchaud}, \cite{ethos} and \cite{flocks}). Their development in quantitative sociology is ongoing \cite{galam}, \cite{durlauf}, \cite{daurlauf2}, \cite{brock}, \cite{montanari}, \cite{loreto}, and they have recently been used to study immigration phenomena  \cite{contucci2}, \cite{BarCon1}.

Following the available data, our research focuses on classical quantifiers of integration such as the fraction of temporary and permanent labor contracts given to immigrants, the fraction of marriages with spouses of mixed origin (native and immigrant), and the fraction of newborns with parents of mixed origin. We aim at a predictive theory by which the magnitude of the above-mentioned indicators can be expressed as a function of the density of immigrants $\gamma$, i.e. the ratio between the number of immigrants $N_{imm}$ and the total population $N=N_{imm}+N_{nat}$, where $N_{nat}$ is the number of natives
\begin{equation}
\gamma \; = \; \frac{N_{imm}}{N_{imm}+N_{nat}} \; \in \; [0,1].
\end{equation}
Since these quantifiers are the sum of many random variables divided by their total number, and we are interested in studying their dependence on $\gamma$, what we seek is first the empirical law from real data and then the theoretical probability law that, in the limit of large numbers \cite{feller},  entails the observed collective behavior.

Within this work we are interested on the average values of the quantifiers, at the national scale. The quantifier fluctuations are shortly analysed
only from the heuristic point of view (see the Supplementary Material). The problem of establishing if our mathematical theory predicts correctly
their structure, as well as the study of the data at different scales, will be subjects of future works.

To introduce our approach, let us step back to a well-understood phenomenon of collective particle behavior like those considered in statistical physics. Let say we are interested in the basic matter of discovering whether those particles behave independently or not. In many instances, in particular for ferromagnetic particles, the knowledge of their collective observables, like the magnetization, is not enough to answer the question since we could observe the same magnetization for systems with or without interaction. Nevertheless, being able to control some parameters like the temperature of the system or the strength of the imitation coefficient (coupling constant) among particles or the external magnetic field, makes the problem easily solvable by the classical theory of magnetism \cite{curie,weiss}. When the interaction is negligible the response of the system to a small solicitation is well approximated by a linear function. If instead the imitation is dominating, the system would remain insensitive to solicitations up to a critical threshold and start responding very quickly once that threshold has been exceeded.

We want to point out that the same problem has emerged in a sociological context too, as was lucidly formulated by Max Weber \cite{weber}: {\it Social action is not identical with the similar actions of many persons... Thus, if at the beginning of a shower a number of people on the street put up their umbrellas at the same time, this would not ordinarily be a case of action mutually oriented to that of each other, but rather of all reacting in the same way to the like need of protection from the rain}.

Individuals, indeed, do behave sometimes independently from each other. For instance the McFadden's Nobel prize rewarded solution of the Bay Area Rapid Transit problem \cite{macfadden} is based on a non-interacting system of individuals as pointed out in \cite{gbc} and \cite{jpb}.

Imitative or, more generally, correlated behavior is nevertheless even more frequent. Classical examples are when the actions of others are imitated because they are fashionable, traditional, or lend social distinction \cite{weber}. Others have pointed out that individual decisions are made according to imitation even in situations that were previously not considered \cite{brock,birne}. For example, sociologists have argued that imitation, or adopting the behavior of others, is a frequent and highly rational strategy when the consequences, social as well as personal, of one's actions are difficult to assess \cite{hedstrom}. Therefore {\it ex ante} both types of behavior are candidates when seeking to explain how integration comes about.

Our research shows how it is possible to distinguish, using quantitative methods, whether the value of the integration quantifier follows from people acting according to some individual preferences independently of other people, as in Max Weber's rainfall example, or whether it follows as a result of social interaction with others and, of course, all the possibilities in between. The two extreme cases are described in statistical mechanics either as free theory-- independent particles, perfect gas \cite{thompson} -- or interacting theory with possible phase transitions \cite{mpv}.

While it is easy to envision formal similarities between particle behavior and human behavior, the difficulty in our context is that the analogy needs to be strengthened beyond the formal level. In the social sciences, in fact, there is no natural notion of system temperature or, in other terms, it is not clear how to measure a {\it cost function} for social actions nor what units to use to perform the measurement. Moreover, there is no simple way to tune the strength of imitation between people or that of their individual tendency to decide toward some choice. Our proposal to overcome the obstacle is to consider, as control parameter of the Immigrants-Natives system, the quantity that tunes the total number of available cross-links couplings among the two populations:
\begin{equation}
\label{gumg}N_{imm}N_{nat} \; = \; \Gamma(\gamma) N^2 = \; \gamma (1 - \gamma)N^2 \; ,
\end{equation}
i.e.
\begin{equation}\label{gumg}
\Gamma(\gamma) = \gamma (1 - \gamma) \; .
\end{equation}

\begin{figure}[h]
\centering
\includegraphics[width=0.5\textwidth]{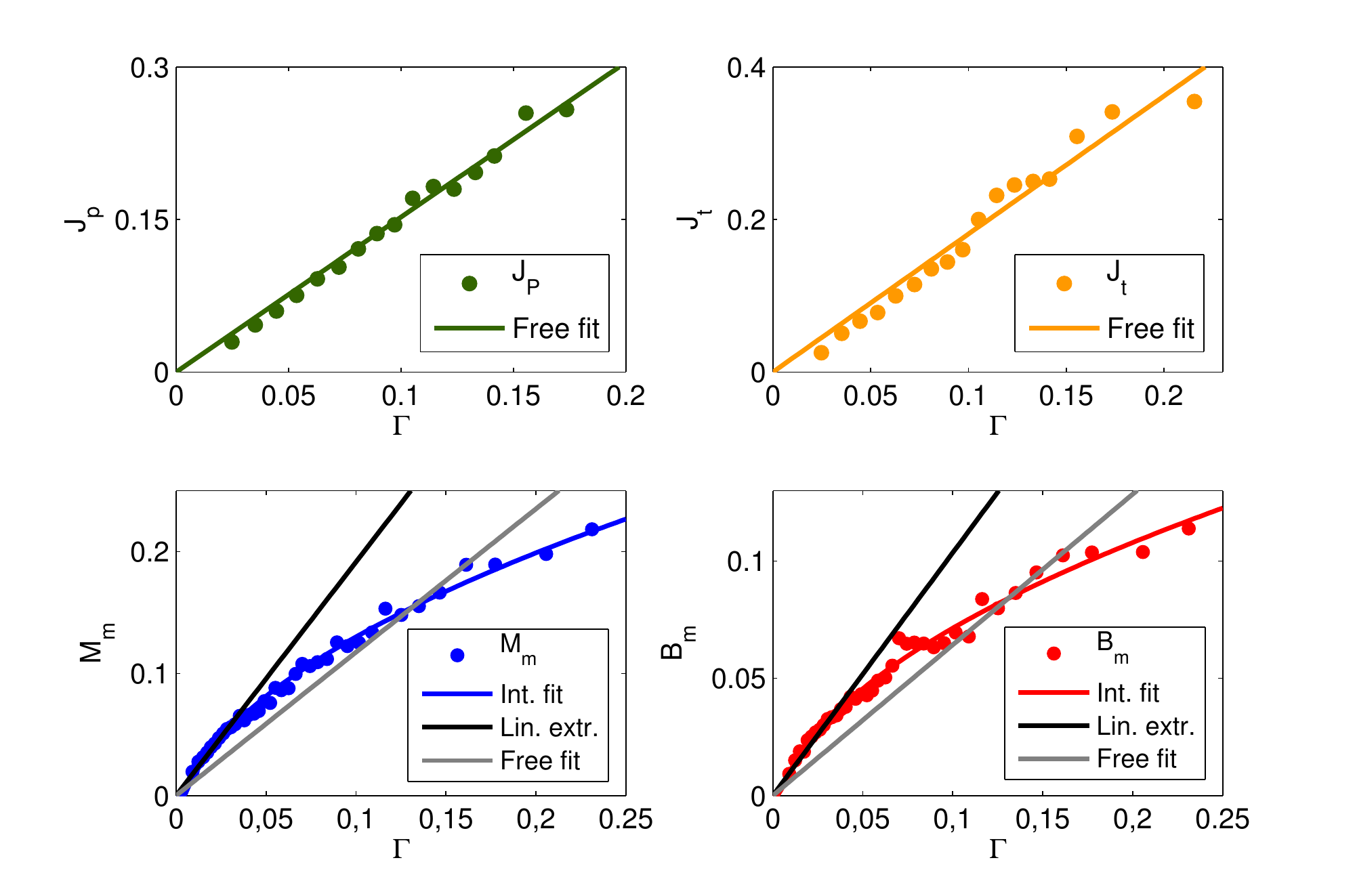}
\caption{\label{sommario} Dots are average quantities versus $\Gamma$.
Left upper panel: quantifier $J_p$ (green dots), the fraction of permanent labor contracts given to immigrants on the total of labor contracts, with the best linear fit (free fit) $a\Gamma$ ($a=1.52\pm0.05$, goodness of fit $R^2=0.985$). Right upper panel: quantifier $J_t$ (yellow dots), fraction of temporary contracts given to immigrants, with the best linear fit (free fit) $a\Gamma$ ($a=1.81\pm0.09$, with a goodness of fit $R^2=0.963$). Left lower panel: quantifier $M_m$ (blue dots), fraction of mixed marriages, with the best square root fit (blue curve) $c\sqrt{\Gamma}$ ($c=0.53\pm0.02$, goodness of fit $R^2=0.992$), the best linear free fit (grey line) $a\Gamma$ ($a=1.18\pm0.07$, with a goodness of fit $R^2=0.855$) and the best linear extrapolation fit (black line)  $b\Gamma$ ($b=1.92\pm0.07$, for $0< \Gamma\le 0.035$, goodness of fit $R^2=0.964$). Right lower panel: the quantifier $B_m$ (red dots) fraction of newborns with mixed parents, with the best square root fit (red curve) $c\sqrt{\Gamma}$ ($c=0.28\pm0.01$, goodness of fit $R^2=0.984$), the best linear free fit (grey line) $a\Gamma$ ($a=0.64\pm0.05$, goodness of fit $R^2=0.789$) and the best linear extrapolation fit (black line)  $b\Gamma$ ($b=1.04\pm0.05$, for $0< \Gamma\le 0.04$, goodness of fit $R^2=0.922$).}
\end{figure}

The main results of our work are summarized in Fig. \ref{sommario} where the average values of the quantifiers are
shown at increasing densities (notice that, in a small neighborhood of zero the parameters $\Gamma$ and $\gamma$ coincide): while the quantifiers measuring labor market integration (green and yellow dots in figure) exhibit linear growth in $\Gamma$, the quantifiers capturing the intensity in mixed-marriages and the number of newborns with mixed parents (blue and red dots) display a non-linear behavior. The distinctive feature of this non-linear growth is
its quick take off which progressively decreases at increasing densities. Hence, if we apply a linear theory when interpreting the whole database (grey line), we would underestimate the level of integration when the immigrant density is low and overestimate it when it is high. Moreover, an integration forecast based on a linear theory (black line) would lead to predictions that are twice as large as the observed values. A finely tuned fit of the blue and red points shows a functional shape proportional to $\sqrt{\Gamma}$ starting very close to the origin of the axes (blue and red curves).
The data analysis calls for a theory able to describe all the observed data on a unified mathematical framework: we proceed by building such theory by first observing that a labor contract, a marriage, a child birth are coupling relations among humans. In a two-group system such as a society composed of immigrants and natives, there can be in-group or cross-group couplings. Consequently, the choice between, say, marrying or hiring an immigrant over a native is dichotomous. To this end the Discrete Choice theory proposed by McFadden \cite{macfadden} is a natural candidate to describe the frequency of cross-group couplings in large populations. McFadden's theory contains a crucial assumption of mutual independence between the involved random variables \cite{gbc}. When applied to the case we are studying, that theory predicts a linear
growth in $\Gamma$ in a suitable interval. Our findings indicate first how well McFadden's theory works in assessing the level of integration in the Spanish labor market and suggest that the choices between assigning the job to a native or to an immigrant are made in a mutually independent fashion, case by case, no matter how other actors choose, with little or no peer-to-peer mechanism at all. Our results show, however, that the choice of marrying or having a child with a native or an immigrant partner is not well described by the classical discrete choice theory.
\begin{figure}%\label{data}%[t]
\centering
\includegraphics[width=0.5\textwidth]{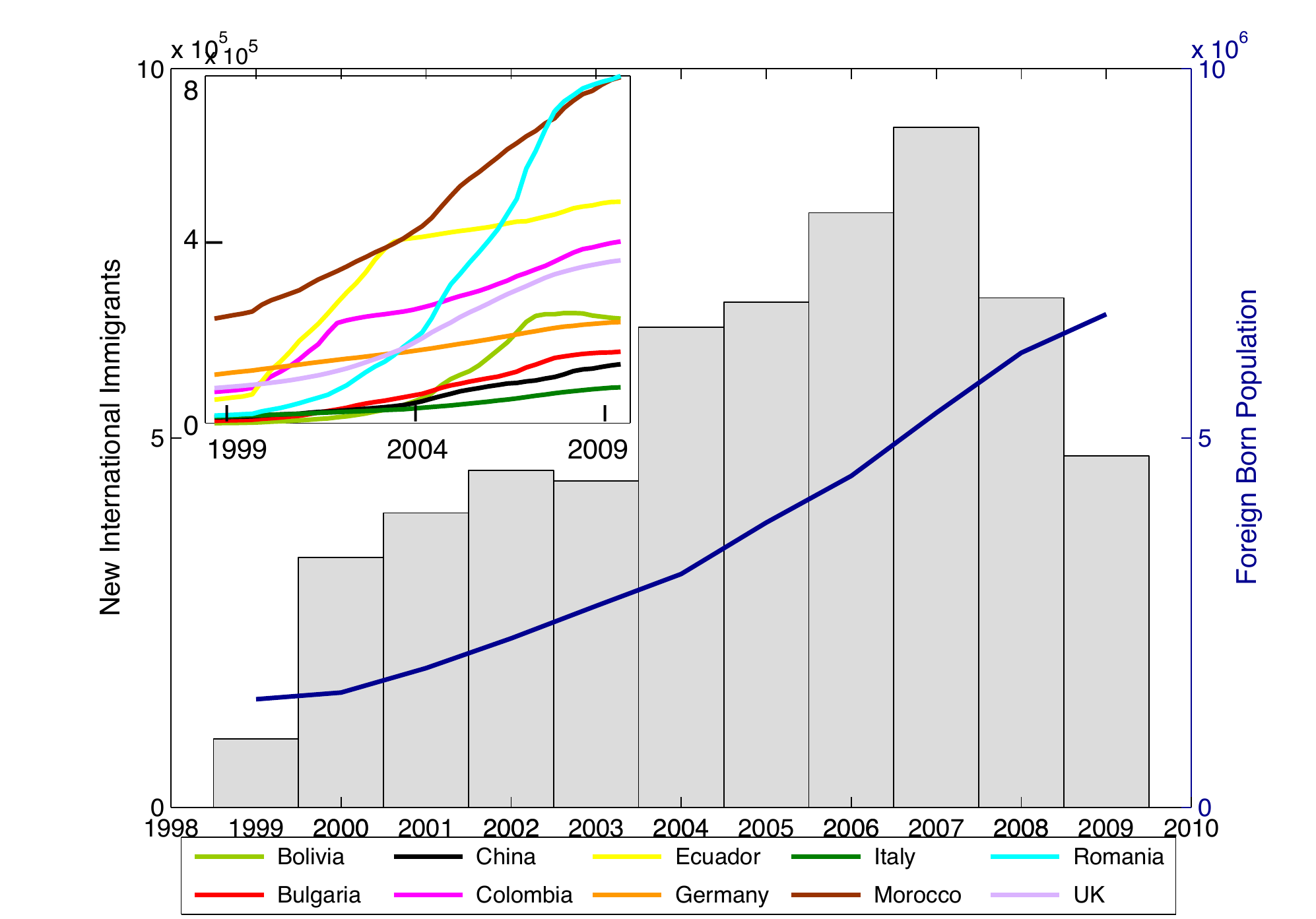}
\caption{\label{datatime} International immigration and stock of foreign born population in Spain in the decade $1999-2009$. The inset highlights migrant income from specific (main) countries of origin.}
\end{figure}
%below is RS editing of this. it needs a bit more work I think
We argue that this mismatch arises because the latter decisions are no longer taken independently. Instead the action of marrying an immigrant or having a child with an immigrant is a decision that is contingent of how others in the environment have acted or not acted in this context before the decision at hand. As a first step, and coherently with the concept of {\em universality} in statistical physics \cite{thompson, mpv}, we do not consider the causes of such social influence (like observing and imitating already successful strategies, or imitate the behavior of people we trust, or comply with cultural norms) but we simply refer to all \textit{social actions} that may cause interdependence or {\it imitation} in decision making.

Theories that relax the assumption of independence and cater to imitation, introduced in \cite{brock} and developed in \cite{gbc,CG1,CG2}, predict a square root behavior of the probability of cross-group couplings as observed in the marriage and child birth data. The methodological part of our work identifies a mathematical model capable of dealing with both situations for different values of the parameters. The result is a generalization of the monomer-dimer model \cite{Liebb1} with the addition of an imitative interacting social network component, with random topology in agreement with the {\it small world}-scenario \cite{smallworld} (see the section Supplementary Material). The model we propose reduces to the classical discrete choice theory with linear growth in situations when imitation is negligible, and to the square root behavior when imitation is prevailing.

The paper is organized as follows. The first section is Data and Integration Quantifiers where we introduce the data and
define the quantities to be studied. In the second section Results and Discussion we present the findings and their theoretical
interpretation, as well as some possible outlooks. Finally,
in the Supplementary Material, articulated in {\em Data Analysis and Elaboration} and {\em Statistical Mechanics Methodology},
we outline details of our experimental and mathematical framework.

%%%

\section*{Data and Integration Quantifiers}

We use immigration data from Spain on the time
interval $1999$ to $2010$ because it corresponds to the period in which Spain received most of its immigrant population. The smallest geographical unit for which data are available is for the administrative unit, Municipality. Hence, in this work Location equals Municipality.

Data on local immigrant densities are compiled as follows. We use the size of the immigrant population and the native population in each municipality as reported in the $2001$ Census as our baseline. Thereafter, we estimate the local immigrant densities for different points in time (quarterly) between 1999 and 2010. The analysis is based on the information contained in the Statistics over residential variation in Spanish municipalities, the so called Estadistica de variaciones residenciales (EVR): for each municipality, the data contain information about (internal) migration to and from other Spanish municipalities, as well as all international migration events and statistics on vital events (births and deaths), as elaborated by Spain's National Statistical Agency (INE). A unique feature of the Spanish data is that they also include so-called undocumented immigrants, that is, immigrants who lack a residence permit \cite{Sandell}. Undocumented immigrants are usually not included in official statistical sources. However, their share of the immigrant population is often significant, and excluding them would underestimate the true size of the immigrant population and, most likely, change the nature of the studied phenomena. We point out that the data we analyzed didn't include detailed topological features on location to allow us
to address issues like municipality segregation and mix of migrant groups.

Data on marriages and births are drawn from the local offices of Vital Records and Statistics across Spain (Registro Civil), and have been compounded by the INE. Data on marriages contain information about the time of the marriage as well as the place of birth, nationality, municipality of residence, and the like, of all the spouses entering into marriage in Spain. By our definition, a mixed marriage occurs when a Spanish-born (native) person marries a person born in a foreign country. Similarly, data on births contain information about the place of birth, nationality, municipality of residence, among other things, of all the newborns parents. For the same reasons as with mixed marriages, we consider all newborns with one native and one foreign born parent to be newborns with parents of mixed origin. Information is included on mixed marriages and newborns with parents of mixed origin where the foreign-born spouse or parent is an undocumented migrant. We focus our analysis on birth and marriage events that occurred during the period 1999 to 2008. However, data on density, marriages and births are subject to minor data protection restrictions. An individual residence municipality is only disclosed if its population is larger than $10,000$. For this reason, out of approximately $8,000$ municipalities in Spain, our analysis focuses on only 735. Still, $85\%$ of Spanish immigrants reside in the included municipalities.

Data on labor contracts come from Spain's Continuous Sample of Employment Histories (the so called Muestra Continua de Vidas Laborales or MCVL). It is an administrative data set with longitudinal information for a $4\%$ non-stratified random sample of the population who are affiliated with Spain's Social Security. Sampling is conducted on a yearly basis. We use data from the waves 2005 to 2010. The inclusion of an individual in the sample is determined by a sequence in the
individual's social security number that does not vary across sample waves. This means that individuals are maintained across samples. New affiliates with a social security number matching the predetermined sequence are added in each new wave. The data contain information on contractual conditions such as whether the individuals have a temporary or indefinite labor contract, as well as the contracts start and stop times. Residential data at the level of municipality and information about place of birth are also available. In contrast to the data on densities, marriages, and births, for these data the residence municipality is only disclosed if the population is larger than $40,000$.

For those unfamiliar with the Spanish immigration context, the following brief information may be useful. In 1999, Spain received fewer than $50,000$ new documented and undocumented immigrants. Since then, annual immigration levels have increased dramatically, reaching a peak in 2006 and 2007, with inflows exceeding $800,000$ (see light gray bars in Fig. \ref{datatime}). Spain's documented and undocumented foreign born population has risen from little more than 1 million to over 6.5 millions in the analyzed period (see solid line in Fig. \ref{datatime}). Its share of the total population has risen from less than $3\%$ to over $13\%$ in the same period. Currently there are immigrants from almost all nations in Spain. However, some 20 immigrant origins account for approximately $80\%$ of Spain's total immigrant population. Immigrants from Romania form the largest minority in Spain ($767,000$ at the end of 2008), followed by immigrants from Morocco ($737,000$ at the end of 2008) and Ecuador ($479,000$ at the end of 2008). Europe and South America together account for over $70\%$ of Spain's total immigrant population.

\bigskip

Operatively, we derive two datasets based on the information described above. One contains data on marriages and newborns, and the other on labor market affiliation. Both datasets contain spatial and temporal information, such as the municipal code, quarter, year, and the immigration density in the municipalities across time. The data on labor contracts consist of $3,553$ entries over the period $2005-2010$. The data on marriages and newborns consist of $27,144$ entries spanning the period $1999-2008$. For the overlapping period (16 quarters of the 2005-2008 window), the values of $\gamma$'s match very well, which can be seen as a good test of the quality of the second sample, since the first dataset is not a sample.

The quantifiers we study as a function of $\gamma$ are defined as (we use the symbol $\#$ to mean ``number'') :
\begin{eqnarray}
J_p &=& \frac{\#\ of\ permanent\ contracts\ to\ immigrants }{\#\ of\ permanent\ contracts}, \\
J_t &=& \frac{\#\ of\ temporary\ contracts\ to\ immigrants }{\#\ of\ temporary\ contracts}, \\
M_m &=& \frac{\#\ of\ mixed\ marriages}{\#\ of\ marriages}, \\
B_m &=& \frac{\#\ of\ newborns\ with\ mixed\ parents}{\#\ of\ newborns} \; .
\end{eqnarray}
We notice that they can be studied equivalently in $\Gamma$ by the quadratic map of the interval
$0 \le \gamma \le 1/2$ into $0\le \Gamma \le 1/4$.

\section*{Results and Discussion}

\begin{figure}%[t]
\centering
\includegraphics[width=0.5\textwidth]{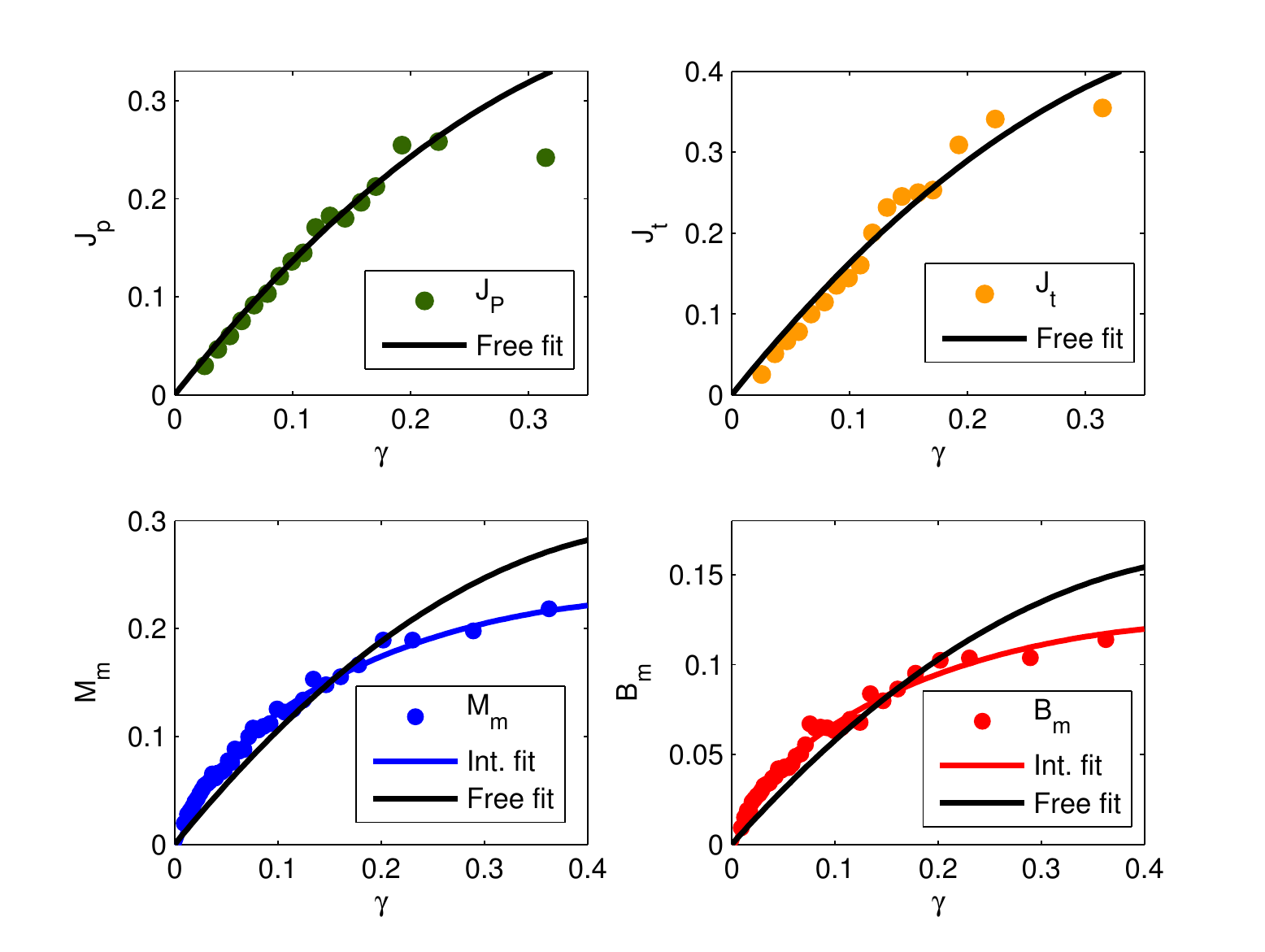}
\caption{\label{main}Upper-left panel, $J_{p}(\gamma)$: Data are represented as spots (green) in each bin. The black line is the best fit of the free theory outcome $J_{p}(\gamma) = c_F \gamma (1-\gamma)$ and yields the best value $c_F=1.52 \pm 0.05$ with a coefficient of determination $R^2 = 0.985$. Upper-right panel: $J_{t}(\gamma)$: Data are represented as spots (yellow) in each bin. The black line is the fit of the free theory outcome with $J_{t}(\gamma) = c_F \gamma (1-\gamma)$ and yields the best value $c_F = 1.81 \pm 0.09$ with a coefficient of determination $R^2 = 0.963$. Lower-left panel, $M_m(\gamma)$: Data are represented as spots (blue) in each bin. The black line is the best fit of the free theory outcome $M_m(\gamma) = c_F \gamma (1-\gamma)$ and yields the best value $c_F=1.18 \pm 0.07$ with a coefficient of determination $R^2 = 0.855$. The blue line is the fit of the interacting theory outcome with $M_m(\gamma) = c_I \sqrt{\gamma(1-\gamma)}\Theta(\gamma-\gamma_c)$ with $c_I= 0.53 \pm 0.02$, $\gamma_c=0.0047\in (0.0034, 0.0064)$ and with a coefficient of determination $R^2 = 0.992$. Lower-right panel, $B_m(\gamma)$: Data are represented as spots (red) in each bin. The black line is the fit of the free theory outcome with $B_m(\gamma) = c_F \gamma (1-\gamma)$ and yields the best value $c_F = 0.64 \pm 0.05$ with a coefficient of determination $R^2 = 0.789$. The red line is the fit of the interacting theory outcome with $B_m(\gamma) = c_I \sqrt{\gamma(1-\gamma)}\Theta(\gamma-\gamma_c)$ with $c_I=0.28 \pm 0.01$, $\gamma_c=0.0036\in (0.0022,0.0057)$ and with a coefficient of determination $R^2 = 0.984$.}
\end{figure}

The four classical integration quantifiers just introduced describe some type of social coupling like the one between the employer and the employee in the job market, or between individuals in a marriage or parenthood. The reader can find in the Supplementary Material all the details to obtain the average points from the raw data, as well as the fitting procedure used to obtain the functional relation on $\Gamma$ from the averaged data as
shown in Fig. \ref{sommario}. For completeness we also show the quantifiers behavior versus the original parameter $\gamma$ in Fig. \ref{main}.

We now provide some statistical physics theoretical bases to explain the observed functional relations. The mathematical details
of their derivation are given in the Supplementary Material.

Let us first discuss the job market. We start by observing that the number of employers is proportional to the number of natives. This can be seen by computing, in the period of time we study, the correlation between the size of the native
population and the total number of employers from the data made available online by the Spanish Bank ``La Caixa" (Anuario Economico de Espa\~na), which turns out to be
$0.98$. On the other hand, the number of immigrant employees is proportional to $\gamma$. An elementary combinatorial computation
(as well as a probabilistic one based on tossing Bernoulli coins \cite{feller}) predicts a frequency of jobs given to immigrants of the following type
\begin{equation}\label{libera}
P = c_F \gamma(1-\gamma)= c_F \Gamma(\gamma) ,
\end{equation}
where $c_F$ is some proportionality constant. This formula, which is linear in $\Gamma(\gamma)$, provides a remarkably good fit to the employment data for both permanent and temporary jobs, as one can see from the upper panels of Figures \ref{sommario} and \ref{main}. By giving each job position a two-valued random variable, since the job is given either to an immigrant or to a resident, one sees that the main feature of models predicting a similar behavior in $\gamma$ is the assumption of the mutual {\it independence} of those random variables. In other words, on the macroscopic scale we are working on (i.e. the whole country), the likelihood of giving a job to an immigrant is indifferent to the fact that another job has been given to an immigrant or not. Such models within socioeconomic sciences, are all versions of the {\it discrete choice theory} \cite{macfadden, gbc} originally devised to predict the use of public transportation but nowadays also used in other contexts such as occupations, residency locations, etcetera. That method, by suitably measuring the parameters of the theory (by polls or historical series), can yield quite accurate predictions as it did in the Bay Area Rapid Transit problem \cite{macfadden}. They are usually parameterized according to the logit (or multi-logit) probability distribution. Their predictive success is based on the fulfillment of the independence hypothesis. Potential threats for its validity are then peer-to-peer effects, belief propagation effects, and in general all the situations where individual rational choice is perceived as difficult and people instead resort to imitation or anti-imitation of others. Within statistical mechanics, a theory of independent random variables is also called a {\it free} theory as opposed to an {\it interacting} one. The discrete choice theory is quite well-suited for policy-making for several reasons. First, it is based on empirical data and second, the utility function it is built on allows researchers to test concrete policy scenarios by varying the free parameters.
%\textbf{(it may be worth stressing further that fits with the logit in this framework are completely equivalent to fitting with the hyperbolic tangent that stem from the statistical mechanics route, as ultimately, both the functions encode the binary choices available to the decision makers)}.

Coming to the other two quantifiers, the mixed marriages and newborns, one can see that the same type of free theory that provided an excellent fit for the data in the job market case is much less appropriate here (see Fig.\ref{sommario}). Indeed, the data show an anomalously high growth at small values of $\gamma$, largely underestimated by the free theory, followed by a crossover, after which the free theory overestimates the quantifier. The free theory makes a mistake by up to $30\%$ of the quantifier.

The first thing to check is whether the different nature of the random variables can be responsible for the difference in social behavior. Social couplings, for instance in marriage, do fulfill the interaction rule of monogamy. That is, each individual is married only to one other individual, unlike employers who may have many employee or an employee who may hold more than one job. Similarly, the children of mixed couples do not on average exceed small units. All these are indeed forms of interaction, but a straightforward computation shows that, if no other interactions are
also present, the same type of behavior of equation (\ref{libera}) still emerges for the probability of mixed marriages and newborns as discussed
in the {\em Statistical mechanics methodology} section of the Supplementary Material.

From the lower panels of Figures \ref{sommario} and \ref{main}, a different type of curve provides an impressively good fit, which is the square
root of the main quantity $\gamma(1-\gamma)$ i.e.
\begin{equation}\label{nosdomine}
P=c_I \sqrt{\gamma(1-\gamma)}= c_I \sqrt{\Gamma(\gamma)},
\end{equation}
for a suitable proportionality constant $c_I$. It is a well known fact that the square-root curve $\sqrt{\Gamma-\Gamma_c}$ carries the
fingerprint of the {\it mean-field} ferromagnetic theory of statistical mechanics \cite{curie}\cite{weiss}
that describes the imitative behavior of particles in dichotomic states. As explained in the Supplementary Material
a vanishing $\Gamma_c$ is compatible with a class of diluted networks that we consider in the proposed mathematical model.
The match between theory and phenomenological data, beside
being visually manifest, is quantitatively represented by the coefficients of determination reported in the captions of Fig. \ref{main}.
We point out moreover that the interacting square-root behavior is clearly emerging from the data even without detailed
information on the topology of the interaction network and proves the robustness of our findings. Neglecting the segregation
or sub-community effects, possibly revealed by detailed topological information, may in fact lead, at worst, to underestimate
the interaction effects.

A natural extension of our work would be to analyze
the fluctuations around the average values. This has been done in the Supplementary Material only at a simple
descriptive level. The main point still to be addressed is a quantitative match between fluctuation data and mathematical theory
beyond their order of magnitude which is, roughly speaking, compatible with the noise generated by finite volume
effects and the randomness of the social network structure. Moreover one should seek for more detailed data
carrying also topological information in order to address segregation effects, as well as an increased detail beyond the bi-partite
scheme immigrant-native to allow subsets of different nationalities and multipartite modeling. A natural test for our
model is to analyse data from different countries. We believe that although the presence or the lack of the interacting regime
may be different in different countries due to differences in cultures and regulations, the model proposed is general enough
to adapt to different cases. In particular, the square root starting from the origin of the axes is somehow the expected emergence behaviour
for the interacting cases. We plan to return on all those matters in future works.

To summarize our work we have analyzed a specific dataset of integration quantifiers in Spain and identified the empirical laws
at growing immigration densities. Focusing on their average values on the national scale we found two types of growth and we
have provided a simple theoretical framework for their interpretation. Our results could improve our ability to target integration
policies since they provide an operative method to distinguish whether a macro phenomenon such as immigrant integration
is the product of {\it social action}, as in the case of intermarriages and newborns with mixed parents, or the product of the
{\it common action} of many people \cite{weber}, as in the labor market case. Our study shows the potential gain in introducing
new families of mathematical models based on a statistical mechanics extension of discrete choice theory, since the latter offers
a set of formal tools to systematically analyze and quantify socioeconomic situations.

\section*{Supplementary Material}
\vskip .2cm

This appendix is used to explain the technical specifications and computational details for both the data analysis and elaboration (first section) and the
theoretical mathematical part on statistical mechanics (second section).

\subsection*{Data Analysis and Elaboration}
\begin{figure}%[h]
\centering
\includegraphics[width=0.5\textwidth]{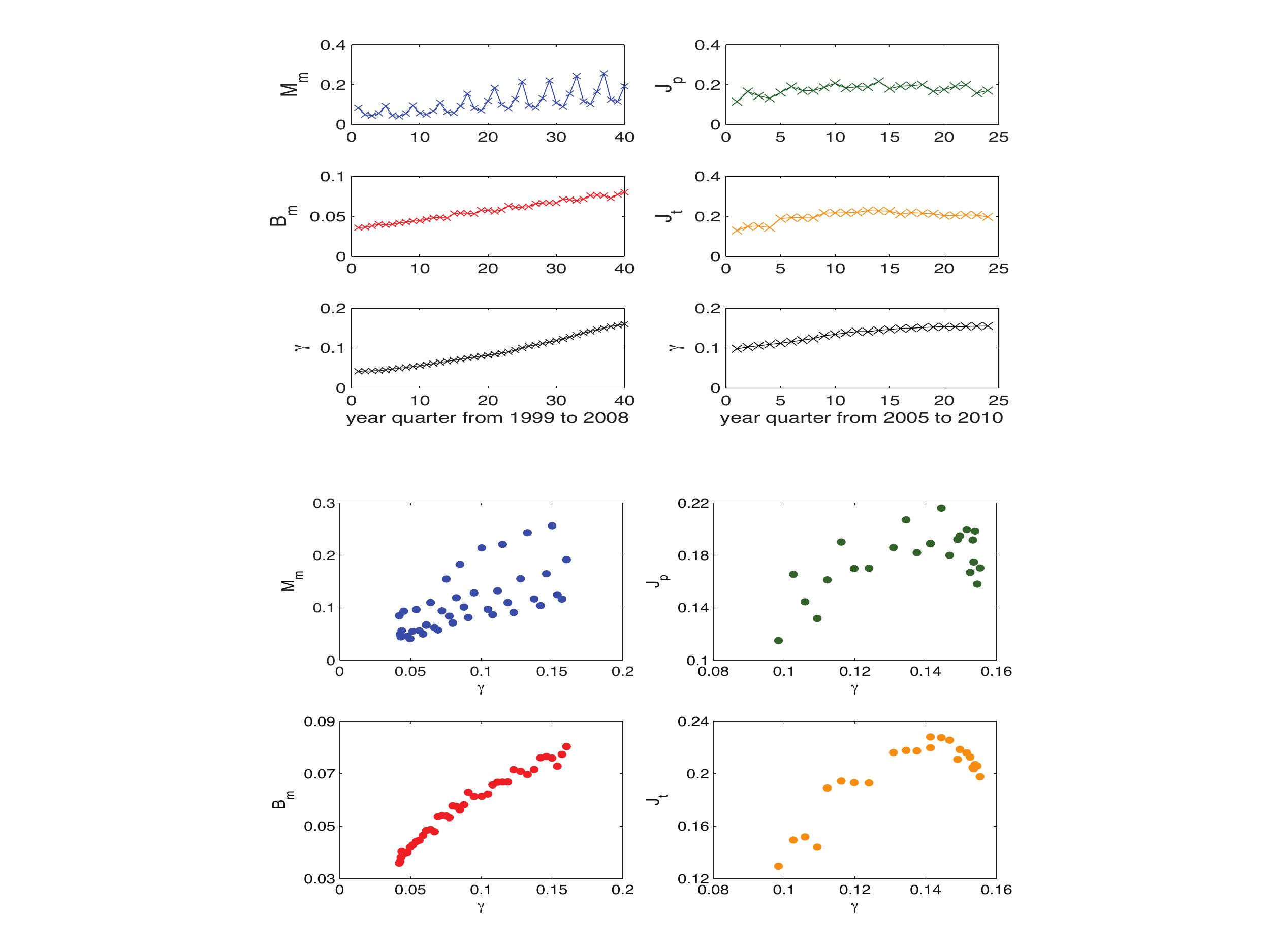}
\put(-127,190){(a)}
\put(-127,99){(b)}
\caption{\label{timeinv} Panel (a): time series representing the quantifiers $\mathbf{Q}$ (mixed marriages $M_m$, newborns from mixed parents $B_m$, permanent jobs $J_p$, temporary jobs $J_t$) and migrant's densities $\gamma$ versus year quarters $\mathbf{t}$ in the two databases. Each point in the plot is the average value of $Q$ in the quarter $t$. Panel (b): quantifiers $\mathbf{Q} $ versus $\gamma$ obtained from time series of panel (a), i. e. $Q(t(\gamma))$, where $t(\gamma)$ is the inverse of $\gamma(t)$ (two lower panels of (a))}.
\end{figure}
To investigate the functional dependence of the quantifiers on $\gamma$, we first tested the time series of the parameters involved.
Starting from the raw data, we consider the average values of each quantifier $Q$ and of the migrant's density $\gamma$ in the two databases as a function of the quarters $t$. The result is shown in  Supplementary Fig. S\ref{timeinv}a.
%Figure \ref{timeinv}a shows how the average quantifiers and migrant's density %behave in the two databases in terms of the quarters.
The newborns with mixed parents display a very regular linear increase over time. The mixed marriages behave similarly but have an added seasonal periodicity as expected from cultural behavior. The two labor quantifiers exhibit a more complex behavior over time. The two lower panels of Supplementary Fig. S\ref{timeinv}a show how the density of immigration increases over time in the two databases. Using those functions $\gamma(t)$ and inverting them in $t(\gamma)$, we can plot each quantifier $Q(t)$ in terms of $\gamma$, thereby obtaining $Q(t(\gamma))$ represented in the four panels of Supplementary Fig. S\ref{timeinv}b.
\begin{figure}%\label{data}%[h]
\centering
\includegraphics[width=0.5\textwidth]{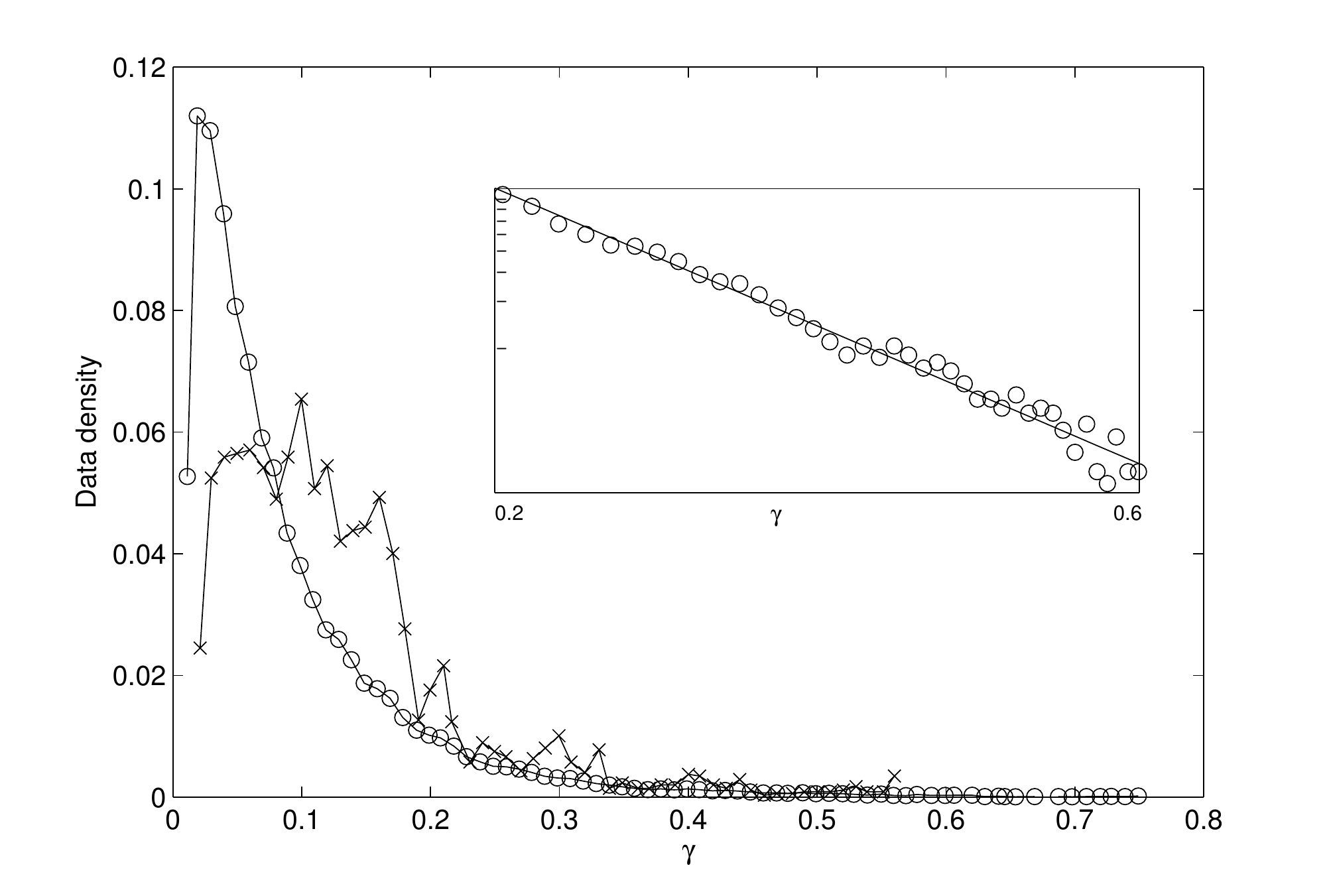}
\caption{\label{datadensity} Density of the marriage and newborn dataset (circles) and of the job market dataset (crosses) as a function of $\gamma$. In the inset the marriage and newborn data density is fitted, for $\gamma\ge 0.2$, with the power-law behavior (in log-log scale) where $\mu(\gamma) \propto \gamma^{\delta}$, $\delta=-3.241 \pm 0.024$.}
\end{figure}
As we can see, apart from a vague functional dependence on the newborns, all of the other quantifiers display erratic behavior and escape a functional law. It is evident that the time fluctuations through which these graphs are obtained contain spurious external effects. To explain better this point a parallel with thermodynamics is very enlightening: the derivation of the law that relates magnetization and temperature from time series would be extremely difficult in a condition in which energy is pumped into a ferromagnet from a random source. In fact we should first derive the laws of magnetization with time and temperature with time. Then we should eliminate the time variable from those data. Large fluctuation would make the data useless. Moreover when those fluctuations seem to be absent (the newborns case), the two processes of marginalization over time and inversion yield a very poor output. The bottom line is that the time series approach is not the suitable method for obtaining the functional dependence we are looking for since it loses relevant information and propagates spurious external effects.

To fully use the rich information of the two databases and extract from them the functional dependence of the quantifiers in terms of $\gamma$, we first proceed by identifying the empirical probability distribution ensemble for each dataset. We do so by merging into a unique catalogue the data entries in each database, regardless of their coordinates in space and time, and ordering them by increasing values of $\gamma$. The observed time windows cover a time scale much larger than the typical time scales involved in the dynamics of the jobs market or marriages/newborns that we focused on. Analysis of their density versus $\gamma$ (Supplementary Fig. S\ref{datadensity}) shows that, for both datasets, only about one percent of the data are found for $\gamma \geq 0.4$. To efficiently model the macroscopic behavior of the integration quantifiers with robust statistics, we limit our study below that threshold. We also notice that data density decreases for small $\gamma$. The reason for this is that our observation window started when the migration phenomena was already running and the density of migrants in Spain was larger than zero. The immigrant densities appear to have tails with power law distribution. For the marriages and newborn dataset, as highlighted in the inset, we find for $\gamma \ge 0.2$ the law $\mu(\gamma) \sim \gamma^{\delta}$ with $\delta=-3.241$.
\begin{figure}%[h]
\centering
\includegraphics[width=0.5\textwidth]{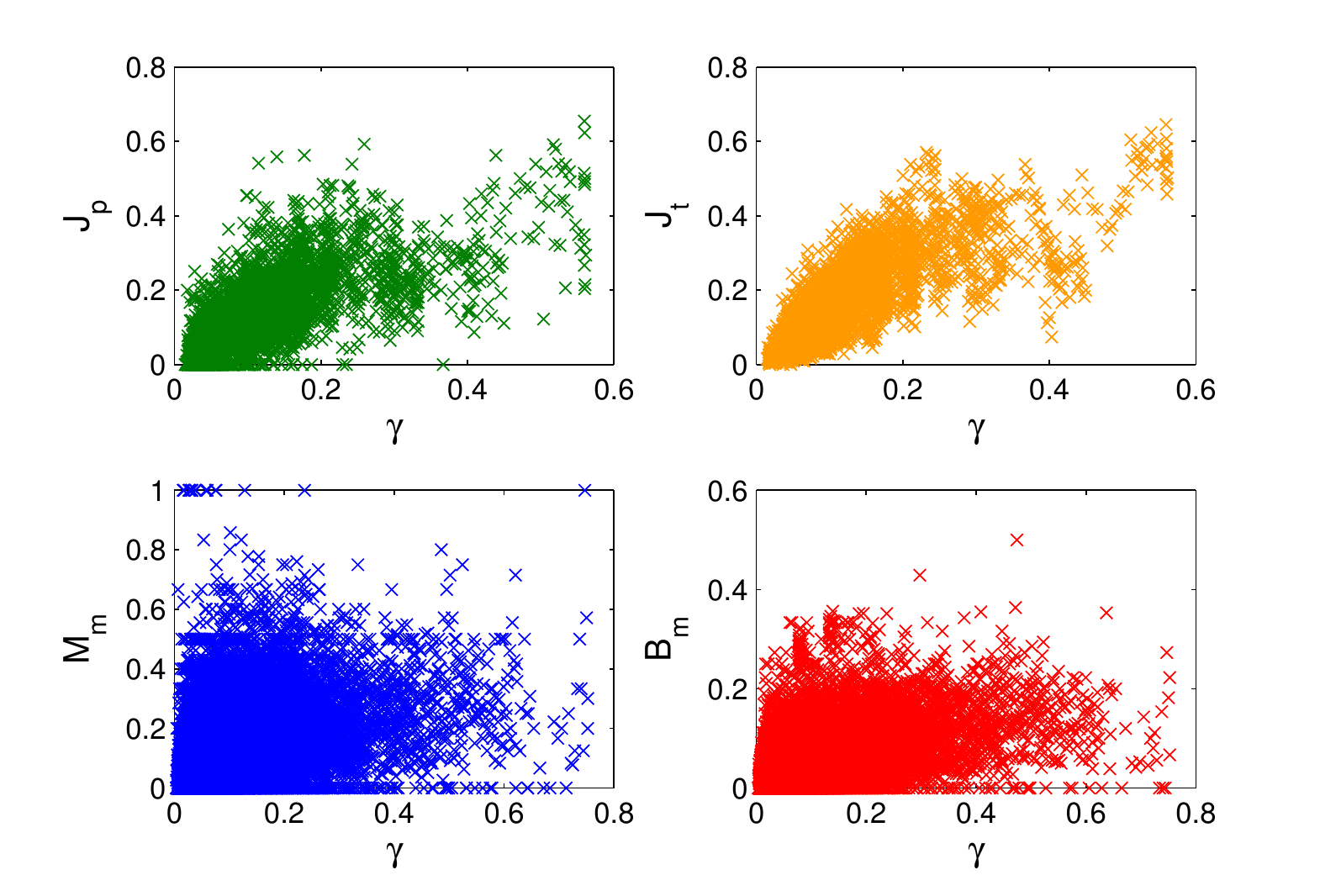}
\caption{\label{nuvola} Raw data versus $\gamma$.
Green points represent the fraction of permanent job positions held by migrants in a municipality where a percentage $\gamma$ of migrants is present (apart from restrictions outlined in the introduction, the whole of Spain is sampled over the entire analyzed timeframe); similarly orange points account for temporary jobs. Further, blue points represent the fraction of mixed marriages, while red ones mirror the newborns from mixed parents. One may note that data in the left panel seem to lie along horizontal lines displaced according to $1/n$, with $n\in\mathbb{N}$ due to seasonal preferences in weddings. See the discussion within the paper.}
\end{figure}
\begin{figure}%[h]
\centering
\includegraphics[width=0.5\textwidth]{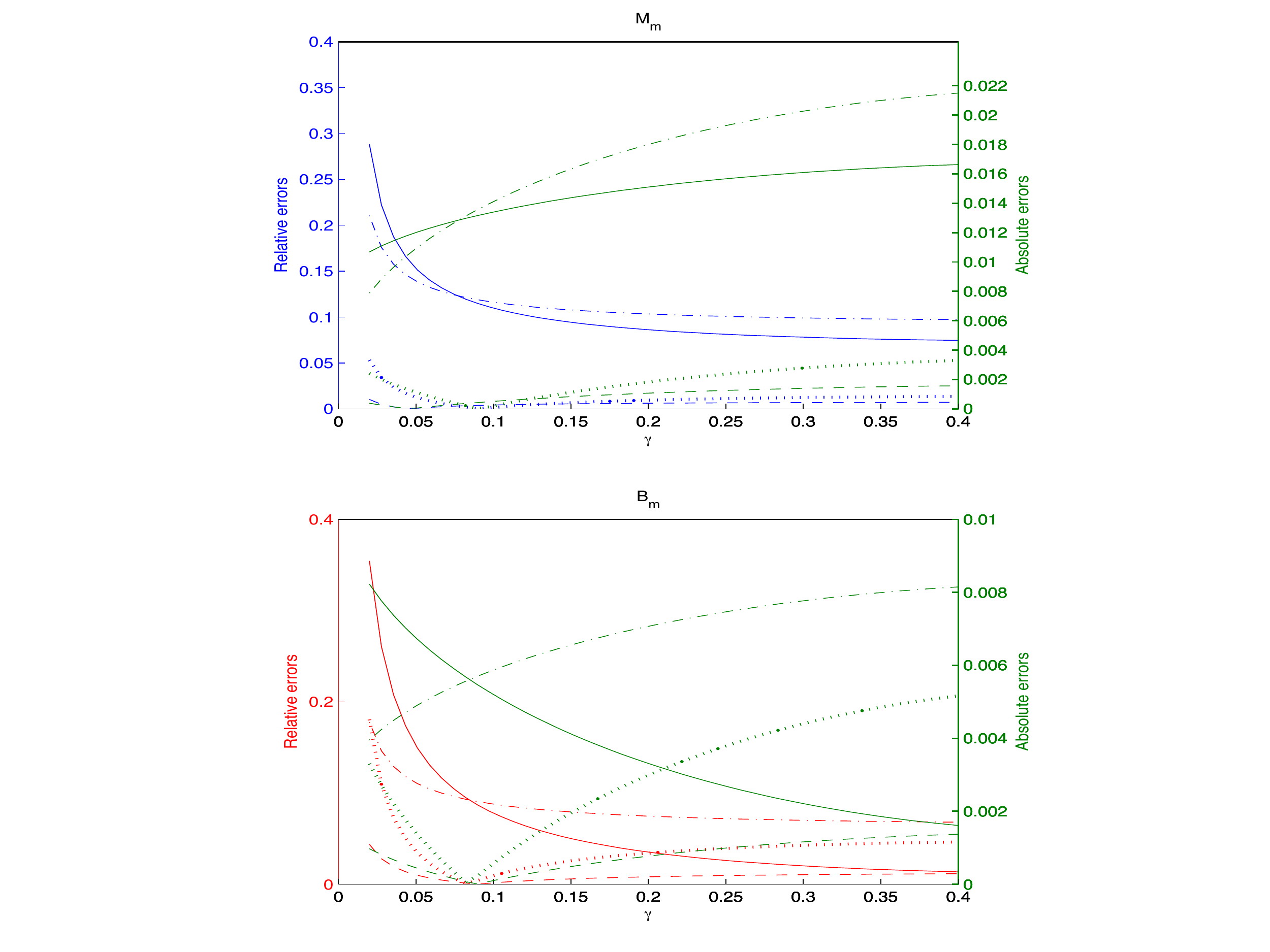}
\caption{\label{errori} Upper panel: Relative errors $|Q_i-Q_j|/Q_j$, $i\ne j$ $i,j\in$ \{A,B,C,D\} (blue lines) and absolute
errors $|Q_i-Q_j|$, $i\ne j$ $i,j\in$ \{A,B,C,D\} (green lines) as a function of $\gamma$ for
$Q=M_m(\gamma)$. Lower panel: Relative errors $|Q_i-Q_j|/Q_j$, $i\ne j$ $i,j\in$ \{A,B,C,D\} (red lines) and absolute errors $|Q_i-Q_j|$, $i\ne j$ $i,j\in$ \{A,B,C,D\} (green lines) as a function of $\gamma$ for
$Q=B_m(\gamma)$.
$Q_A$ denotes the mean value using binning with constant information, $Q_B$ the mean value using binning with constant width, $Q_C$ the mediant with constant information binning and $Q_D$ the mediant with constant width binning.
For both panels the continuous lines refer to $i=B, j=D$  and the dash-dotted lines to $i=A, j=C$ and represent the errors made using the mean approach with respect to the mediant one with constant step binning and with constant information binning, respectively. The dotted lines ($i=D, j=C$) and the dashed lines ($i=B, j=A$) represent the errors made using the constant step binning with respect to the constant information, with the mean approach and with the mediant approach, respectively.}
\end{figure}
Supplementary Figure S\ref{nuvola} shows the raw data {\it clouds} for each quantifier.
An apparent anomaly is the presence in the lower left panel of horizontal lines where the data agglomerate. A further analysis shows that their values are due to the fractions with small denominators, that is, municipalities where the total number of marriages within the observed quarter does not exceed the few units. The explanation of this anomaly is found in the strong cyclical behavior due to seasonal preferences about the appropriate time for marriage in Spain (see also Supplementary Fig. S\ref{timeinv}). People in Spain prefer to marry in the summer rather than in the winter, which is unsurprising. Aggregating the data from quarters to years would wash away the anomaly. Nevertheless, as we explain below, it turns out that roughening the data in this way is not necessary, and it is possible to keep the dataset as it is, and hence preserve its richness.

Since we are interested in the quantifier's averages as functions of $\gamma$ and since all quantifiers are ratios, there are two possible ways of computing the averages. For a given bin of $\gamma$ one can compute 1) the statistical average of the ratios, or 2) the ratio between the statistical average of numerators and the statistical average of the denominators. The first is the usual mean of the ratios and the second is their global {\it mediant}. As will be explained later the difference in the results obtained from the two distinct procedures is in the range of $0.1$ to $0.2$ percent (see Supplementary Fig. S\ref{errori}) and they can consequently be considered as effectively equivalent.

We then proceed by grouping the data into bins over $\gamma$ in which the averages can be evaluated.
We found that $15$ to $20$ bins optimizes the job market dataset, while $35$ to $40$ optimizes the dataset on marriages and newborns. For the binning criteria, we tested the method of constant information and that of constant bin width. The advantage of the first criterion is a constant robustness quality across all bins. Clearly, with this approach, the width of the bin will vary over $\gamma$ (their width increases at high values of $\gamma$ due to the data density decrease reported in Supplementary Fig. S\ref{datadensity}). This can be avoided if we instead use constant bin width. However, with this approach the tradeoff is that as $\gamma$ increases, the amount of data inside each bin may diminish (in particular for $\gamma \geq 0.4$). As we confined our analysis to $\gamma_{max}=0.4$ for robustness requirements, the two criteria produce essentially the same results as one can see
in Supplementary Fig. S\ref{errori}.

%Figure \ref{main}
Figure 3 (upper panels) in the main text shows the outcome of the average criteria and coarse graining procedure for $J_p$ and $J_t$. The
dot's plots are made of $17$ bins. On each bin the dot represents the average value of $200$ data. On each figure the black curve is the {\it free fit}, that is, the curve of type $c_F\gamma(1-\gamma)$ best fitting the experimental points. Their goodness of fit, reported in the relative captions, is estimated as $R^2_{J_p(\gamma)}\sim 0.985$
and  $R^2_{J_t(\gamma)}\sim 0.963$.

%Figure \ref{main}
Figure 3 (lower panels) in the main text shows the results for $M_m$ and $B_m$. In this case, the plots of the dots are made of $38$ bins and each of them comes from $700$ points. In this case, the free fit has a much lower goodness of fit: $R^2_{M_m(\gamma)}\sim 0.855$ and $R^2_{B_m(\gamma)} \sim 0.789$. In particular, the data show an anomalously high growth rate for small $\gamma$ and a low one for large $\gamma$. For this reason we tested another family of curves, whose genesis we aim to explain through statistical mechanics in the next sections, and ultimately account for interactions among persons. Remarkably, all of these curves scale as $c_I \sqrt{\gamma(1-\gamma)}\theta(\gamma-\gamma_c)$ where $\theta$ is the step function. The agreement of the fit with the experimental data is clearly shown by the values $R^2_{M_m(\gamma)} \sim 0.992$ and $R^2_{B_m(\gamma)} \sim 0.984$.
In the above formula, one may note the classical exponent {\it one half} typical of theories that account for imitative interaction. The presence (or lack) of a critical value $\gamma_c$ is determined by the underlying social network. It is known that in ferromagnetic theories network dilution eventually decreases $\gamma_c$ to zero, without affecting the critical exponent for a large class of social topologies \cite{smallworld}\cite{AglBar1}\cite{BarAgl2}\cite{SW1}\cite{barra5}\cite{vespignani}\cite{barabasi}\cite{stanley}\cite{ginestra}. Accordingly, we found empirical values $\gamma_c \sim 10^{-3}$ as reported in the caption.

%For each quantifier $Q$ of Figure \ref{main} the single dot, representing the average within the bin, can be %obtained using
For each quantifier $Q$ of Figure 3 the single dot, representing the average within the bin, can be obtained using
two averaging process (mean and mediant) and two kind of binning criteria for $\gamma$ (bins with constant information and with constant width). In Supplementary Fig. S\ref{errori} we test the robustness of our findings against these possible choices.
Therefore, for each quantifier $Q$ in the $\gamma$-bin we compute $Q_A$, the mean value using binning with constant information, $Q_B$ the mean value using binning with constant width, $Q_C$ the mediant with constant information binning and $Q_D$ the mediant with constant width binning. In Supplementary Fig. S\ref{errori} we represent the relative errors ($|Q_i-Q_j|/Q_j$, $i\ne j$ $i,j\in$ \{A,B,C,D\} ) and absolute errors ($|Q_i-Q_j|$, $i\ne j$ $i,j\in$ \{A,B,C,D\})  as a function of $\gamma$ made by comparing in couples all this possible choices for mixed marriages and newborns from mixed couples.
Note that the relative errors are expected to increase at high values of $\gamma$, while in that region they are reduced with respect to those for smaller $\gamma$ by at least one order of magnitude in all observables. The apparent increase of the relative error sizes at small $\gamma$ is due instead to a ratio between small numbers and is a simple and harmless consequence of numerical noise as confirmed by the behaviors of their absolute values.

For completeness we also analyse the quantifier fluctuations from the averages, whose results are shown in Supplementary Fig. S\ref{f5}. To obtain those distributions, we proceed as follows. Let us focus on a concrete example, e.g. the permanent jobs reported in the main text in Fig. 1
%\ref{sommario}
(green plot) and the corresponding quantifier's fluctuations shown in Supplementary Fig. S\ref{f5} (green plot). We take each dot in the first picture (representing the average of $J_p$ within its corresponding $\Delta \Gamma$ bin) to be the center of the (green) distribution reported in Supplementary Fig. S\ref{f5}. Of course, within the chosen $\Delta \Gamma$ bin of Fig. 1,
%\ref{sommario},
not all the data values contributing to obtain the average will be exactly the value of the mean (represented by the green spot), but some one will be smaller (and fall on the right in the corresponding panel of Supplementary Fig. S\ref{f5}) and some other will be bigger (so to fall on the left). Summing on all the deviations from the corresponding means of each spot in the same (green) plot of Fig. 1,
%\ref{sommario},
when all the points have been investigated, the green histogram of Supplementary Fig. S\ref{f5} is then finished and we move to another quantifier.
\newline
Still in Supplementary Fig. S\ref{f5}, beyond the reported bare results, the latter are also fitted with standard distribution laws as Gaussian, Logistic, Gumbel, and Cauchy (the goodness of fits are reported in the captions). The rescaled distribution for $M_m(\gamma)$ is slightly asymmetric to the right: the Gumbel distribution (compared with Logistic and Gauss distribution) gives the best fit. The rescaled distribution for $J_p(\gamma)$, $J_t(\gamma)$ and $B_m(\gamma)$ is more symmetric: in these cases the Logistic distribution (compared with Cauchy and Gauss distribution) yields the best fit. In the insets the normal probability plots for the rescaled quantifiers show that the tails of the sample distributions are non-Gaussian.
\begin{figure}%[h]
\centering
\includegraphics[width=0.5\textwidth]{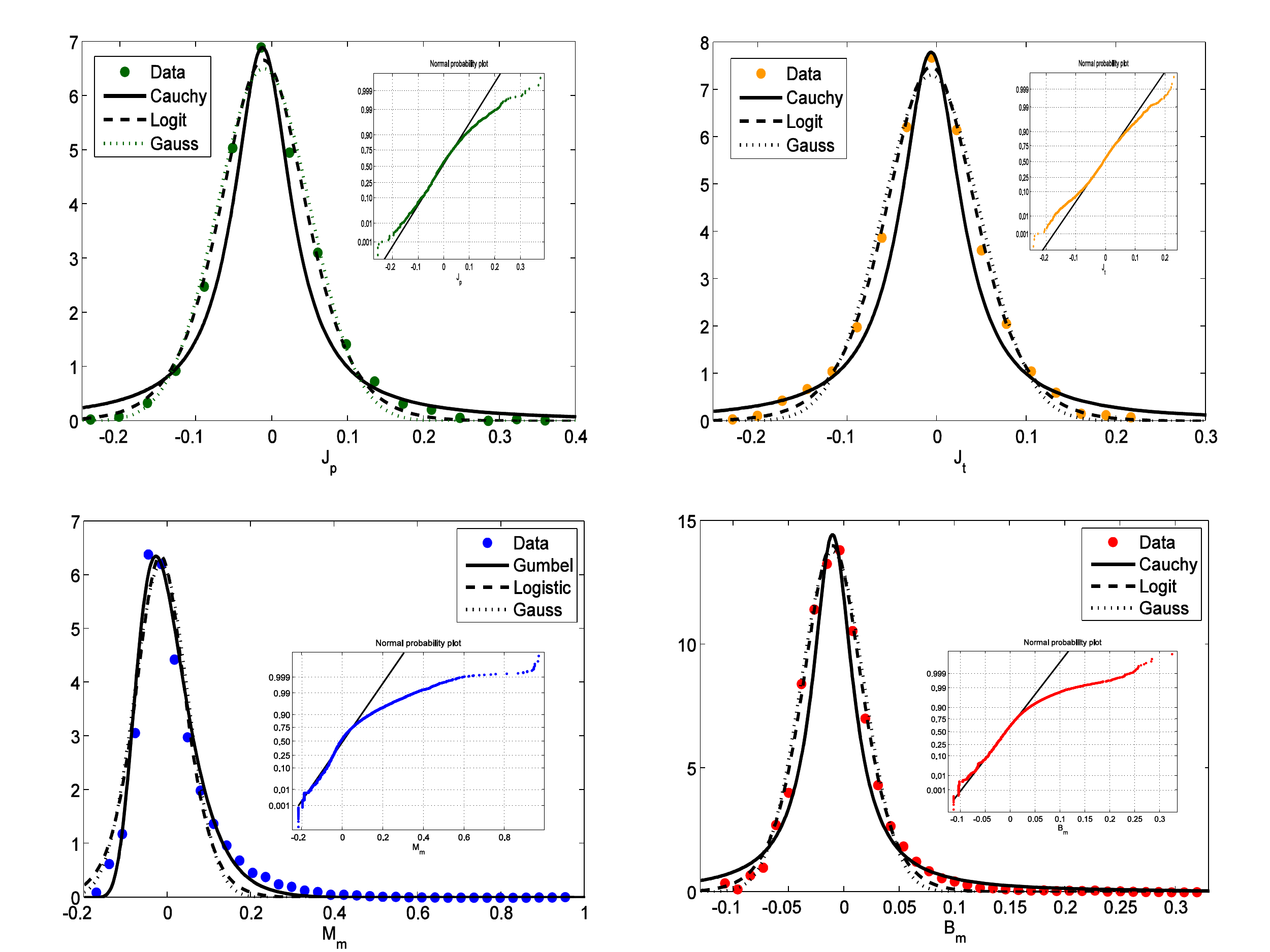}
\caption{\label{f5} Left upper panel: rescaled distribution of $J_p(\gamma)$.
Fit with standard distribution laws are reported: Cauchy distribution $R^2 = 0.951$, Logistic distribution $R^2 = 0.994$ and Gaussian distribution $R^2 = 0.986$.
Right upper panel: rescaled distribution of $J_t(\gamma)$.
Fit with standard distribution laws are reported: Cauchy distribution $R^2 = 0.956$, Logistic distribution $R^2 = 0.996$ and Gaussian distribution $R^2 = 0.984$.
Left lower panel: rescaled distribution of $M_m(\gamma)$.
Fit with standard distribution laws are reported: Gumbel distribution $R^2 = 0.979$, Logistic distribution $R^2 = 0.956$ and Gaussian distribution $R^2 = 0.941$.
Right lower panel: rescaled distribution of $B_m(\gamma)$.
Fit with standard distribution laws are reported: Cauchy distribution $R^2 = 0.958$, Logistic distribution $R^2 = 0.994$ and Gaussian distribution $R^2 = 0.988$.
For all panels the inset represents the normal probability plot.}
\end{figure}
\subsection*{Statistical Mechanics Methodology}

This section is devoted to Methods, in particular to fill the gap between the empirical laws and the theoretical apparatus of statistical mechanics,
especially devoted to the readers who are closer to the hard sciences. Our aim is the introduction of a mathematical model based on individual choices, with an internal structure (microscopic theory) whose emergent social behavior (macroscopic theory) reproduces all the observed quantifiers. The model reduces to the standard discrete choice theory \cite{macfadden} when peer-to-peer interaction is negligible and to its interacting version when imitation is instead the dominating factor.

% We start by characterizing pictorially the possible configurations of the two phenomena of mixed marriages and newborns
% in Fig. \ref{example}, then we focus on the details of the statistical mechanics formalism.

To introduce the statistical mechanics model, we assign to each person their own tendency to marry versus remaining single. In addition, each couple $(i,j)$ has their own likelihood to marry or not. Similarly, each person has an individual tendency to have children, and each couple too. All of these phenomena are then described by individual random variables and couple random variables.

The two observables about marriages and children are of course different: for marriages, the monogamy law only allows each individual to belong to a single couple. Newborn couplings instead may not only have multiplicity but individual may have children with different partners regardless of being married or not.

All of these rules, from the mathematical point of view, turn into topological constraints in the configurational space like the {\it hard-core} interaction of monogamy, or probabilistic constraints such as the concentration of children per couple around small integers. The rule structure can be described as follows: given a set of points $1,...,N$, a configuration of marriages $M$ is a set of links among the $N$ points with the property that no points belong to more than one link.
%(see left panel in Fig. \ref{example})
We indicate the unpaired individuals (singles) by $S_M$ and the paired ones (married couples) by $C_M$.
In principle, one should account for possible different numbers between the two genders; however, such refinements are small corrections of the main theory due to the heterogeneous origin of migrants, which balances polarized incomes from countries with eradicated cultural constraints.
We call the set of marriage configurations ${\cal M}$. We want to describe a system in which we assign to each configuration a statistical weight and a partition function built on individual random variables as well as on couple random variables. Calling $s_i$ the weight of the person $i$ in the single state and $c_{i,j}$ that of the couple $(i,j)$ in the married state (both the $c$'s and the
$s$'s are positive real numbers), the partition function (grancanonical) of the system is given by
\begin{equation}
Z^{(\cal M)} \; = \; \sum_{M\in {\cal M}} \prod_{(i,j)\in C_M}\epsilon_{i,j}c_{i,j}\prod_{i\in S_M}s_i \; ,
\end{equation}
where the numbers $\epsilon_{i,j}\in \{0,1\}$ are the acquaintance matrix elements of the population, i.e. they specify the reciprocal knowledge among two individuals. It may be worth stressing that since no topological insights were available within our data we lack specific information on
$\epsilon_{i,j}$. For this reason we carried our analysis with the diluted mean-field approach. While a fully connected mean-field approach would set a critical value $\Gamma_c > 0$, the assumption of an (over-percolated) underlying topology (e.g. a random graph a' la Erdos-Renyi \cite{AglBar1} or a small world a' la Strogats-Watts \cite{smallworld}), implies a rescaling of the critical value with the size $N$ \cite{vespignani}\cite{ginestra}\cite{dembo}, hence $\Gamma_c \to 0$, in full agreement with our experimental findings.
\newline
Similarly a configuration of filiations $F$
%(see right panel in Fig. \ref{example})
is a set of links among the $N$ points with the property that for a given couple $(i,j)$ (not necessarily married) the number of children (links) is distributed according to (say) a Poisson distribution $\rho$ of given average $\lambda$. The choice of the Poisson distribution is the most reasonable, but our conclusions are robust, pathological cases apart. We indicate individuals without children by $U_F$ (undescended) and the couples with children by $P_F$ (parents). We call the set of filiations ${\cal F}$. Calling $u_i$ the weight of the person $i$ in the undescended state and $p_{i,j}$ that of the child $(i,j)$ in the parental state (both the $u$'s and the $p$'s are positive real numbers) the partition function of the system is then given by
\begin{equation}
Z^{(\cal F)} \; = \; \sum_{F\in {\cal F}} \rho(F) \prod_{(i,j)\in P_F}\epsilon_{i,j}p_{i,j}\prod_{i\in U_F}u_i \; .
\end{equation}
%In treatable physics models \cite{BarAgl2}\cite{ginestra}, the $\epsilon_{i,j}$ account for the topology of the system, like d-dimensional lattices, or the complete graph of N-points or more refined structures like Erdos-Renyi and small world graphs.
The random variables ($c,s,p,u$) are taken to be constant on mean field models like the one we treat explicitly in this work. Calling $K_M$ the total number of links in the configuration $M$ and defining the frequency as $\nu_M=K_M/(N/2)$, the expected value of the marriage frequency can be computed as
\begin{equation}
P_{\cal M} \; = \; {\rm Av}\frac{\sum_{M\in {\cal M}}\nu_M \prod_{(i,j)\in C_M}\epsilon_{i,j}c_{i,j}\prod_{i\in S_M}s_i}
{\sum_{M\in {\cal M}} \prod_{(i,j)\in C_M}\epsilon_{i,j}c_{i,j}\prod_{i\in S_M}s_i} \; ,
\end{equation}
where the average operation $Av$ is computed on the acquaintance matrix ensemble. Similarly for the newborn problem, calling $K_F$ the total number of links in the configuration $F$ and defining the fraction $\nu_F=K_F/(N\lambda/2)$, its expected value is
\begin{equation}
P_{\cal F} \; = \; {\rm Av}\frac{\sum_{F\in {\cal F}}\nu_F \rho(F) \prod_{(i,j)\in P_F}\epsilon_{i,j}p_{i,j}\prod_{i\in U_F}u_i}
{\sum_{F\in {\cal F}} \rho(F) \prod_{(i,j)\in P_F}\epsilon_{i,j}p_{i,j}\prod_{i\in U_F}u_i} \; .
\end{equation}
For each population, the previous probability measure provides an average value of the two observables. \\
We now turn to the theory of bi-populated systems where $s_i$ and $u_i$ take two values each, depending only on the individual being on $Imm$ or $Nat$ and the couple variables ($c_{i,j}$ and $p_{i,j}$) take only three values for the three cases $(Imm, Imm)$,
$(Nat, Imm)$ and $(Nat, Nat)$.
We may include an imitative ($J\geq 0$) interaction between the two populations with the introduction of a suitable mean-field Hamiltonian \cite{AglBar1}\cite{BarAgl2}\cite{CG1}\cite{CG2}
\begin{equation}
H(M)=-J_M\sum_{i\in Nat, j\in Imm}\epsilon_{i,j}\sigma_i\sigma_j \; ,
\end{equation}
where
\begin{equation}\label{fasi}
\sigma_i=\begin{cases} +1 &\mbox{if $i$ belongs to a mixed marriage} \\
-1 &\mbox{otherwise }, \end{cases}
\end{equation}
and a similar definition for $H(F)$:
\begin{equation}
H(F)=-J_F\sum_{i\in Nat, j\in Imm}\epsilon_{i,j}\tau_i\tau_j \; ,
\end{equation}
where
\begin{equation}\label{fasi}
\tau_i=\begin{cases} +1 &\mbox{if $i$ has a child within a mixed couple} \\
-1 &\mbox{otherwise }. \end{cases}
\end{equation}
Note that $\sigma$'s and $\tau$'s configurations are uniquely determined by monomer-dimer configurations.
We point out that the case where imitation and anti-imitation coexist \cite{mpv} leads to a different scenario
(see \cite{BarCon1} for a case study).
The two complete partition functions are then:
\begin{equation}
Z^{(\cal M)} \; = \; \sum_{M\in {\cal M}} e^{-H({M})} \prod_{(i,j)\in C_M}\epsilon_{i,j}c_{i,j}\prod_{i\in S_M}s_i \; ,
\end{equation}
\begin{equation}
Z^{(\cal F)} \; = \; \sum_{F\in {\cal F}} e^{-H({F})} \rho(F)\prod_{(i,j)\in P_F}\epsilon_{i,j}p_{i,j}\prod_{i\in U_F}u_i \; .
\end{equation}
We point out that the introduction of an exponential Hamiltonian deformation of the monomer-dimer model is
a working hypothesis to be tested against experimental data and it has the same significance of the
logit distribution assumption in the original McFadden discrete choice theory. For a discussion
about its paramount justification in terms of Entropy variational principles see \cite{flocks} and
references therein.

Calling $M_M$ the number of mixed marriages in the configuration $M$, and defining the {\it frequency of mixed marriages} $f_M=M_M/K_M$, we have that its expected value, that is, the probability of mixed marriages is
\begin{equation}\label{20}\small
P_{\cal M}^{(Nat,Imm)}={\rm Av}\frac{\sum_{M\in {\cal M}} f_M e^{-H({M})} \prod_{(i,j)\in C_M}\epsilon_{i,j}c_{i,j}\prod_{i\in S_M}s_i}
{\sum_{M\in {\cal M}} e^{-H({M})} \prod_{(i,j)\in C_M}\epsilon_{i,j}c_{i,j}\prod_{i\in S_M}s_i}
\end{equation}
and analogously the probability of mixed children
\begin{equation}\label{21}\small
P_{\cal F}^{(Nat,Imm)}={\rm Av}\frac{\sum_{F\in {\cal F}} f_F e^{-H({F})} \rho(F)\prod_{(i,j)\in P_F}\epsilon_{i,j}p_{i,j}\prod_{i\in U_F}u_i}
{\sum_{F\in {\cal F}} e^{-H({F})} \rho(F)\prod_{(i,j)\in P_F}\epsilon_{i,j}p_{i,j}\prod_{i\in U_F}u_i}
\end{equation}
is given in terms of {\it frequency of children from mixed couples} $f_F=M_F/K_F$ where $M_F$
is the number of children from mixed couples in the configuration $F$. Although an exact solution
for the general model introduced in this section is not yet available, one can still obtain results for a
wide variety of cases that include the mono-populated and bi-populated mean field limits
\cite{BarCon1}\cite{CG1}\cite{CG2}. The latter shows two regimes according to the ratio of $J$
(the coupling $J_M$ and $J_F$ tuning the strength of the imitative behavior encoded in the Hamiltonians
$(14)$ and $(16)$, hereafter called $J$ for simplicity) and the monomer-dimer pressure $p=\ln Z/N$. Given
the lack of phase transitions in the Monomer-Dimer model (see \cite{Liebb1,Liebb2} for a rigorous proof
of the non-random mean-field case and \cite{albcont} for the random case), we can focus on the
extreme regimes: the  imitative regime $J\gg p$
in which the interaction $J$ dominates on the Monomer-Dimer interaction (hard core or Poisson), and the free regime
$J\ll p$ where the Monomer-Dimer interaction dominates on the imitative one. The structural difference between
them is the presence of some divergence of the derivative of the $P$'s in formulas (\ref{20}) and (\ref{21})
\begin{equation}
\frac{\partial}{\partial\gamma}P(\gamma) \; .
\end{equation}
In the free regime one finds for the $P$'s a $\gamma$ dependence of the type
\begin{equation}\label{free}
P(\gamma) \; = \;
c_F\gamma(1-\gamma) \; ,
\end{equation}
where the constant $c_F$ depends only on the a priori probabilities of the Monomer-Dimer interaction. The expression can also be obtained from purely probabilistic (or combinatorial) reasoning and always displays a finite derivative in the origin.

In the imitation regime, on the other hand, the interactions among agents encoded in the Hamiltonians favor the imitative behavior, while the Monomer-Dimer interaction term, accounted for by the adjacency matrix of the reciprocal relation, plays the role of a phase-selecting perturbation (the $+$ phase of the Hamiltonians) similar to a small magnetic field in spin models. The diluted mean-field Hamiltonian we introduced has a size proportional to $\gamma(1-\gamma)$, and for various adjacency matrices $\epsilon_{i,j}$ defining diluted topologies (i.e., random graphs, small worlds, etc. \cite{barra5}\cite{AglBar1}\cite{ginestra}\cite{zecchina}), the model predicts a zero (or very small) critical $\gamma_c$ and a functional behavior of the type:
\begin{equation}\label{imitation}
P \; = \; c_I [\gamma(1-\gamma)]^{\frac{1}{2}} \; ,
\end{equation}
where the constant $c_I$, as much as the $c_F$ in the free case, depends only on the a priori probabilities of the Monomer-Dimer interaction. The mechanism underlying such behavior is depicted, as far as the (social) network is over-percolated \cite{bollobas}, and the interactions among agents are only imitative, by the mean-field ferromagnet with the critical exponent {\it one half}. Its relevance
in social sciences has been clearly advocated by Durlauf \cite{durlauf}. Mathematically the critical value of $\gamma$ turns out to be very close to zero as a consequence of dilution \cite{AglBar1}, \cite{ginestra}, \cite{zecchina}, \cite{dembo}. Due to the large ensemble of data we analyzed, we can invoke the law of large numbers \cite{feller}, which allows us to compare experimental frequencies reported in the first part of the paper, with probabilities obtained through the statistical mechanics method. The theoretical models illustrated reach the precise functional behavior of equation (\ref{imitation}) in the limit of infinitely many particles (thermodynamic limit).

\begin{acknowledgments}
We thank Elena Agliari, Tiziana Assenza, Marzio Barbagli, Maurizio Delli Gatti, Cristian Giardin\`a, Claudio Giberti, Sandro Graffi, Francesco Guerra, Maurizio Pisati for interesting conversations.
We thank Enrica Santucci for her help at a very early stage of the manuscript and both the Referees for their interesting suggestions to improve the manuscript.
%FA piacere all Editor che ringraziamo i referees cosi: e' la miglior pubblicita per la rivista!;)
\newline
This work is supported by the FIRB grants $RBFR08EKEV$ and $RBFR10N90W$.
\newline
AB is grateful to Sapienza Universit\`a di Roma; PC is grateful to Universit\`a di Bologna for partial financial support.
\end{acknowledgments}

%%%

%\end{article}

\vfill\eject
\newpage

%%%%%%%%%%%%%%%%%%%%%%%%%%%%%%%%
%%   Figure del Paper
%%%%%%%%%%%%%%%%%%%%%%%%%%%%%%%%

%\section*{Paper Figures: 1-3. Supplementary Material Figures: 4-8.}

\end{document}